\documentclass{Interspeech2024}

\interspeechcameraready 
\usepackage[cmyk]{xcolor}
\usepackage{multirow}
\usepackage{float}
\usepackage{booktabs}
\usepackage{cite}
\usepackage{hyperref}
\usepackage{verbatim}
\hypersetup{
  colorlinks=true,    
  linkcolor=black,     
  citecolor=black,    
  urlcolor=black 
}

\newcommand{\astfootnote}[1]{%
  \begingroup
  \renewcommand\thefootnote{\fnsymbol{footnote}}%
  \footnotetext[1]{#1}%
  \endgroup
}
\usepackage[misc]{ifsym}

\title{PicoAudio: Enabling Precise Timestamp  and Frequency Controllability of Audio Events in Text-to-audio Generation}

\name[affiliation={1,2}]{Zeyu}{Xie}
\name[affiliation={1}]{Xuenan}{Xu}
\name[affiliation={2,3}]{Zhizheng}{Wu}
\name[affiliation={1}]{Mengyue}{Wu$^{\ast}$}

\address{
  $^{1}$X-LANCE Lab, Shanghai Jiao Tong University,
$^{2}$Shanghai AI Lab\\ $^{3}$Chinese University of Hong Kong, Shenzhen
}
\email{\{zeyu\_xie, wsntxxn, mengyuewu\}@sjtu.edu.cn, wuzhizheng@cuhk.edu.cn}


\begin{document}

\maketitle

\begin{abstract}
    Recently, audio generation tasks have attracted considerable research interests. Precise temporal controllability is essential to integrate audio generation with real applications. In this work, we propose a temporal controlled audio generation framework, \textbf{PicoAudio}. PicoAudio integrates temporal information to guide audio generation through tailored model design. It leverages data crawling, segmentation, filtering, and simulation of fine-grained temporally-aligned audio-text data. Both subjective and objective evaluations demonstrate that PicoAudio dramantically surpasses current state-of-the-art generation models in terms of timestamp and occurrence frequency controllability. 
    The generated samples are available on the demo website
    \href{https://zeyuxie29.github.io/PicoAudio.github.io}{\textcolor{cyan}{\textit{https://zeyuxie29.github.io/PicoAudio.github.io}}}.
\end{abstract}

\section{Introduction}
\astfootnote{Mengyue Wu is the corresponding author.}
Recently, significant progress has been made in audio generation. With the advancement of diffusion models, we can now synthesize vivid and lifelike audio segments~\cite{kreuk2022audiogen, yang2023diffsound, liu2023audioldm2, huang2023make, ghosal2023text}. A single model can generate universal audio, including speech, sound effects, and music~\cite{vyas2023audiobox, yang2023uniaudio}. Some researchers are focusing on controllability, such as text-based audio editing or style transfer~\cite{liu2023audioldm1, wang2024audit}, scene control for speech and sound effects~\cite{vyas2023audiobox}, attributes-driven generation~\cite{chung2024t, guo2023audio}, and the generation of extended, variable-length spatial music and sound~\cite{ evans2024fast}.

Although existing models can generate sound by following instructions, when using audio generation models in content creation applications, it's important to control timestamps and the occurrence frequencies of acoustic events precisely.  
Existing models overlook the temporal controllability of the \textit{timestamp, interval, duration, occurrence frequency, and relations like overlap or precedence.}
For example, most models struggle to produce sound occurrences accurately when given text inputs like ``dog barks three times" or timestamps such as ``bird chirping during 4-6 seconds". These limitations significantly affect the models' practical use in generating temporally-controllable audio content.

We argue that the missing of precise controllability in existing audio generation models has their root in the following two aspects:
\textbf{First, the deficiency of temporal control is partially due to insufficient temporally-aligned audio-text data.} The commonly utilized audio-text datasets, such as AudioCaps~\cite{kim2019audiocaps} and Clotho~\cite{drossos2020clotho}, emphasize the fidelity of sound event descriptions and the linguistic sophistication of textual content, but they lack annotations pertaining to temporal aspects. In particular, in the largest audio captioning dataset AudioCaps, the phrase ``xx times", indicating frequency, appears only in $1086/56796 (\approx1.9\%)$ annotations.  Moreover, there are scarce annotations regarding timestamps. High-quality temporally-aligned audio-text data is crucial for training temporal controllable models.  The more meticulously annotated the data, the better the models can learn the precise correspondence between audio outputs and temporal textual conditions, thereby achieving finer-grained control.
\textbf{Second, the diffusion model has limited knowledge of timestamp information.} Existing diffusion-based models aim to learn the relationship between text description and audio event in the audio signal. Although the diffusion models can understand the text instructions at the high level, precise controlling information (e.g. ``event-1 at timing-1 ... and event-N at timing-N") is not taken into consideration. This is due the nature of the current design of the diffusion models, which don't take temporal information into consideration.

\begin{figure}[tbp]
  \centering
  \centerline{\includegraphics[width=1\linewidth]{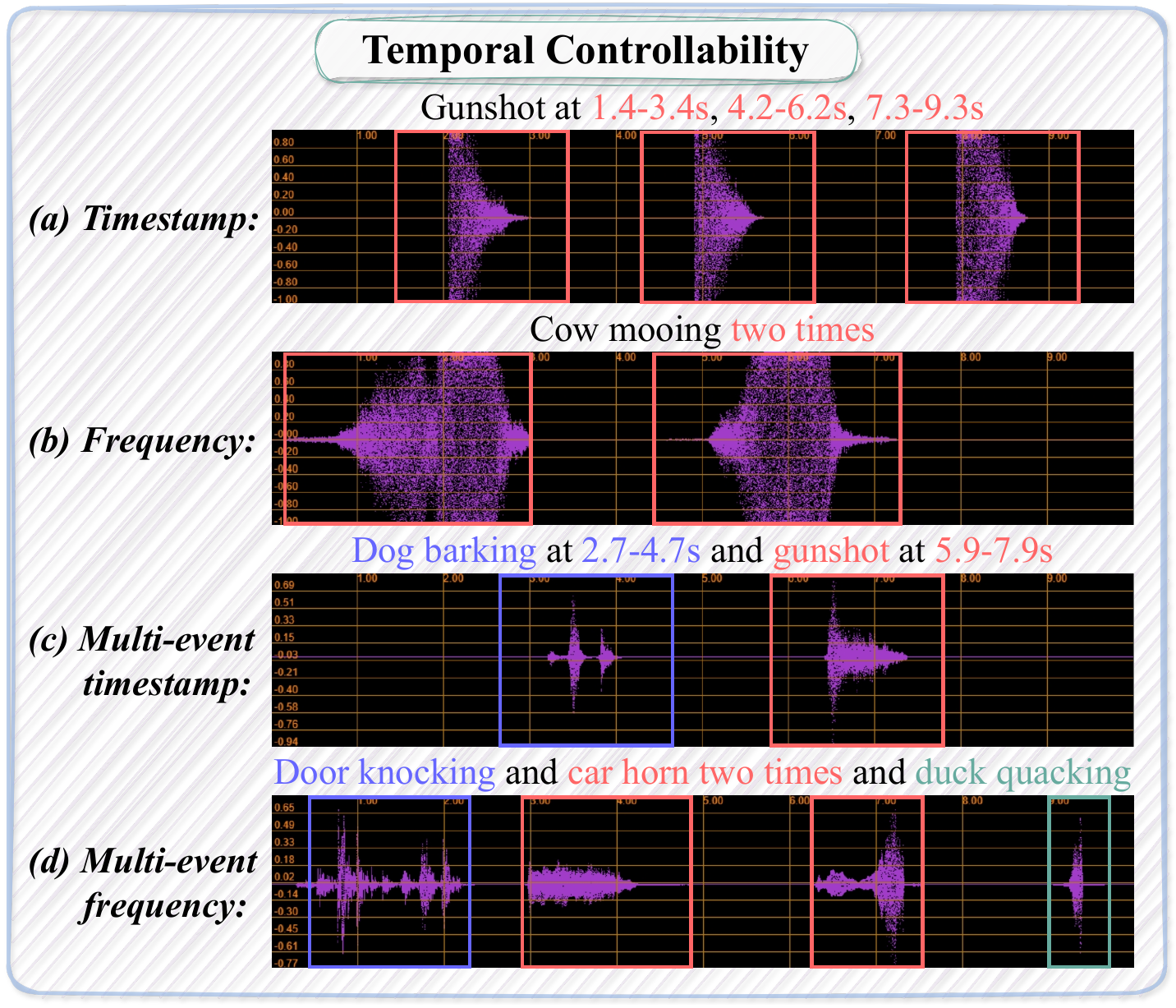}}
  \caption{Illustration of controlling timestamp / occurrence frequency of audio events by PicoAudio. It can enable precise controlling of single events or multiple events.
  }
  \label{fig:intro_example}
 \vspace{-3mm}
\end{figure}

In this work, we propose \textbf{PicoAudio} which enables \textbf{P}recise t\textbf{I}mestamp and frequency \textbf{CO}ntrollability of audio events, by leveraging data simulation\footnote{Simulated datasets for training and evaluation are available at https://github.com/zeyuxie29/PicoAudio}, tailored model designs, and preprocessing with large language model.
We focus on timestamp and frequency control,  while other temporal conditions (e.g., ordering and interval) can be converted into timestamps through textual reasoning, akin to transforming frequency into timestamps in our experiment. PicoAudio proposes a pipeline to simulate data with temporally-aligned annotations.
The pipeline entails crawling data from the Internet, segmenting and filtering audio clips to gather high-quality audio segments, as well as simulating to synthesize realistic audio. PicoAudio introduces tailored modules for temporal control.
(a) Timestamp control is accomplished by incorporating customized input, namely timestamp caption.
With the assistance of large language model (LLM)~\cite{achiam2023gpt}, (b) frequency control, (c) ordering via multi-event timestamp control and (d) multi-event frequency control can be implemented, as shown in Figure~\ref{fig:intro_example}.
Beyond (a)-(d), \textbf{PicoAudio can achieve arbitrary precise temporal control} as long as the LLM is capable of converting the requirement into timestamp captions, which is straightforward for LLM when prompted with simulated data.
Our contributions encompass the following:  
\begin{enumerate}
\item A data simulation pipeline tailored specifically for temporal controllable audio generation frameworks;

\item A timestamp and frequency controllable generation framework, enabling precise control over sound events;

\item Achieving any temporal control by integrating LLM.
\end{enumerate}

\begin{figure*}[htbp]
  \centering
  \centerline{\includegraphics[width=1\textwidth]{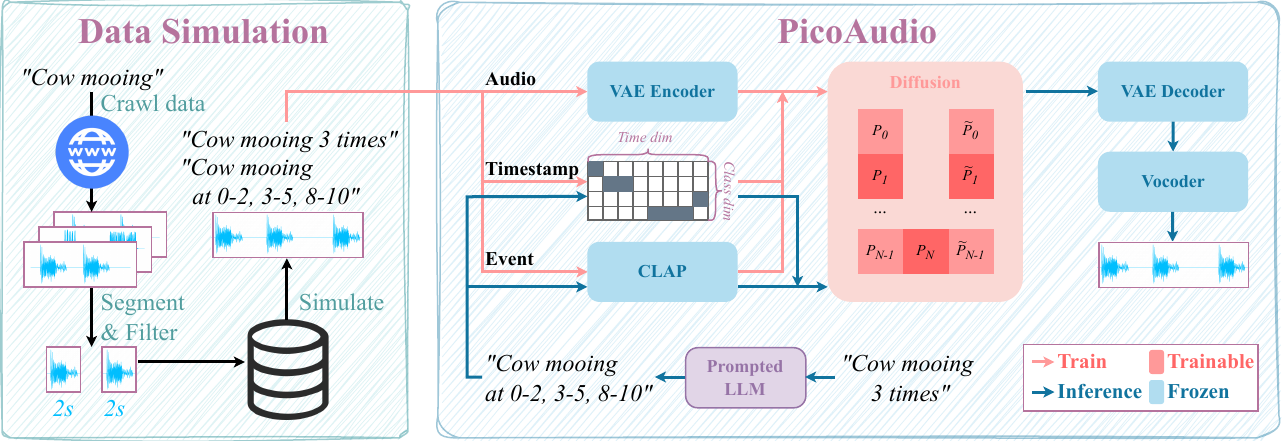}}
    %
    \caption{
    PicoAudio Flowchart. 
    (\textbf{Left}) illustrates the simulation pipeline, wherein data is crawled from the Internet, segmented and filtered, resulting in one-occurrence segments stored in a database.
    \textbf{Pairs of audio, timestamp captions, and frequency captions} are simulated from the database. 
    (\textbf{Right}) showcases the model framework. {\textcolor[RGB]{234,107,102}{Red}} arrows indicate the model training process by using the simulated data. {\textcolor[RGB]{30,120,255}{Blue}} arrows indicate inference based on timestamp or frequency captions, where the LLM is prompted with the simulated training data.
    }
    \label{fig:simulation_model}
\end{figure*}

\section{Temporal Controllable Model}
To enable temporal control in audio generation, we first design a simulation pipeline that automatically acquires data and a tailored text processor to enhance audio generative models' temporal awareness, as shown in Figure~\ref{fig:simulation_model}. 
\subsection{Temporally-aligned Data Simulation}


\paragraph*{Data crawling, segmentation \& filtering}  (1) Audios are crawled from the Internet using event tags as search keywords.
These weakly annotated clips possess only sound event tags and may contain noise.
(2) A text-to-audio grounding model~\cite{xu2021text} is employed to segment crawled data, as it can locate the temporal occurrence of events based on input text. 
Each localized segment encompasses one occurrence of a sound event, such as a ``\textit{2-seconds cow mooing}" segment .
For generality, we also define a burst of continuous short sounds as \textbf{one occurrence}, such as a burst of “keyboard typing” or “door knocking”.
(3) To ensure data quality, a contrastive language-audio pretraining (CLAP) model~\cite{laionclap2023} is utilized for further filtering. 
Thus, we obtain a substantial number of high-quality one-occurrence segments, serving as a one-occurrence database.

\vspace{-3mm}
\paragraph*{Simulation} We randomly select events from the database and synthesize audio by randomly assigning occurrence on-set, following the approach of Xu et al.~\cite{xu2024detailed}. 
The timestamp of occurrence is annotated based on the on-set and the duration recorded in the grounding results.
A simulated pair comprises a synthesized audio and a \textbf{timestamp caption} formatted as \textit{``event-1 at timing-1 ... and event-N at timing-N"}, as well as a \textbf{frequency caption} formatted as \textit{``event-1 j times ... and event-N k times"}.


\subsection{Text Processor} 
The standard format makes rule-based transformations very straightforward.
The one-hot timestamp matrix $\mathcal{O} \in \mathbb{R}^{C \times T}$ is derived from the timestamp caption, where $C$ and $T$ denote the number of sound events and the time dimension, respectively.

\begin{equation}
 \mathcal{O}_{c, t}=\left\{
    \begin{aligned}
    &1, \text{ if event $c$ occurs at time $t$}\\
    &0, \text{ otherwise}\\
    \end{aligned}
\right
.
\end{equation}

LLM demonstrate excellent performance in text processing tasks. 
Thanks to LLM, PicoAudio framework can handle various input formats.
For example, transforming input ``a dog barking occurred between two and three seconds" into the timestamp caption format ``dog barking at 2-3".

LLM also empowers PicoAudio with more capabilities, such as (1) controlling occurrence frequency by transforming ``a dog barks three times" into ``dog barking at 1-2, 3-4, 7-9", and (2) ordering by transforming ``door knocking then door slamming" into ``door knocking at 1-4 and door slamming at 6-8".
The duration of each occurrence is inferred by the LLM based on its own knowledge as well as the examples provided. 
We supplied GPT-4 with $300$ examples in traning set for learning, yielding an initial transformation error rate of $3/1000$ and a refined second transformation error rate of $0/1000$.
It can be observed that the transformation is straightforward for LLM when prompted with simulated training data.

PicoAudio employs a CLAP model~\cite{laionclap2023} to extract event information beyond timestamp, denoted as event embedding $\mathcal{I}$. 
As the timestamp caption also encompass semantic information about sound events, which can also be utilized as guidance. 




\subsection{Audio Representation}
PicoAudio employs a Variational Autoencoder (VAE) for audio representation, given the inherent difficulty in directly generating spectrograms.
The VAE encoder compresses the audio spectrogram 
$\mathcal{A} \in \mathbb{R}^{T\times M}$ into the latent representation $\mathcal{P} \in \mathbb{R}^{T/{2^R}\times M/{2^R}\times D}$, where T, M, R, D denote the sequence length, the number of mel bands, the compression ratio and the latent dimension, respectively.
$\mathcal{P}$ is divided into two halves, representing the mean $\mathcal{P}_{\mu}$ and variance $\mathcal{P}_{\sigma}$
in the latent space. 

The VAE decoder reconstructs the spectrogram $\mathcal{\tilde{A}}$ based on samples from the distribution $\mathcal{\tilde{P}}=\mathcal{P}_{\mu}+\mathcal{P}_{\sigma} \cdot \mathcal{N}(0, 1)$.
The vocoder following the VAE decoder converts the spectrogram back into a waveform.

\subsection{Diffusion}
PicoAudio utilizes a diffusion model to predict 
$\mathcal{\tilde{P}}$
 based on the timestamp matrix 
$\mathcal{O}$ and event embedding $\mathcal{I}$, since
it has demonstrated excellent capabilities in audio generation~\cite{kreuk2022audiogen, yang2023diffsound,liu2023audioldm2, huang2023make, ghosal2023text}. 

The diffusion model encompasses the forward steps that transform representation $\mathcal{P}$ into the Gaussian distribution by noise injection, followed by the reverse steps that progressively denoise.
A noise schedule $\{\beta_n:0 < \beta_n<\beta_{n+1}<1\}$ defines the Markov chain's transition probabilities in the forward steps:
\begin{align}
    q(\mathcal{P}_n|\mathcal{P}_{n-1})\triangleq \mathcal{N}(\sqrt{1-\beta_n}\mathcal{P}_{n-1},\beta_n \mathbf{I})\\
    \mathcal{P}_n=\sqrt{\bar{\alpha}_n}\mathcal{P}_0 + \sqrt{1-\bar{\alpha}_n}\epsilon_n
\end{align}
where $\alpha_n = 1 - \beta_n, \bar{\alpha}_n=\prod_{i=1}^{n}{\alpha_i}$, $\epsilon_n$ follows distribution $\epsilon \sim \mathcal{N}(0,1)$.
At last step $N$, $\mathcal{P}_N $ follows an isotropic Gaussian noise.
The model is trained to estimate noise based on input $\mathcal{O}$, $\mathcal{I}$ and a weight $\lambda_n$ related to Signal-to-Noise Ratio~\cite{hang2023efficient}:
\begin{equation}
  \label{eqn:diffusion_r1}
        \mathcal{L}=\sum_{n=1}^{N}{\lambda_n \mathbb{E}_{\epsilon_n, \mathcal{P}_0}||\epsilon_n - \epsilon_{\theta}([\mathcal{P}_n,\mathcal{O}], \mathcal{I}) ||}
\end{equation}
where $[,]$ denotes concatenation, $\mathcal{I}$ is fused by cross-attention mechanism~\cite{vaswani2017attention}, and
$\epsilon_{\theta}$ denotes the estimation network which can be employed to reconstruct $\mathcal{\tilde{P}}_0$ from $\mathcal{\tilde{P}}_N \sim \mathcal{N}(0,1)$ in the reverse steps with $\bar{  \tau}_n=1-\bar{\alpha}_n$:
\begin{equation}
 \label{eqn:diffusion_r2}
       \mathcal{\tilde{P}}_{n-1}=\frac{1}{\sqrt{\alpha_n}}(\mathcal{\tilde{P}}_{n}-\frac{\beta}{\sqrt{\bar{  \tau}_n}}\epsilon_{\theta}(\mathcal{[\tilde{P}}_n, \mathcal{O}], \mathcal{I}))+\sqrt{\frac{\bar{  \tau}_{n-1}}{\bar{  \tau}_n}\beta}\epsilon
\end{equation}

\section{Experiment}
\subsection{Data Simulation}
\label{sec:simulation}
Audio clips are crawled from Freesound\footnote{https://freesound.org/} using sound event as
search keywords.
Segmentation and filtering are conducted by a text-to-audio grounding model~\cite{xu2024towards} and LAION-CLAP~\cite{laionclap2023} with threshold set to $0.5$ and $0.3$, respectively. 
The collection process results in a total of $636$ high-quality one-occurrence segments containing $18$ sound events.
During simulation, the sound events and on-set time are randomly assigned, with the proportion of $1$, $2$, and $3$ occurrences for each sound event being approximately $2:2:1$.
A total of $5000, 400, 200$ clips are simulated for training, single-event testing and multi-event testing, respectively.

Four temporal control tasks are designed: (a) single-event timestamp control using timestamp caption as input; (b) single-event frequency control using the frequency caption \textit{``xx k times"} as input, which is directly fed into the baseline models.
GPT-4 predicts the duration of segments and subsequently converts frequency captions into timestamp captions before feeding them into PicoAudio.
(c) multi-event timestamp and (d) multi-event frequency control employ captions with multiple events.

\subsection{Experiment Setup}
The time resolution in the timestamp matrix is set to $40$ ms, which implies that temporal control can be achieved with precision at the millisecond level.
The LAION-CLAP~\cite{laionclap2023} is utilized as the event embedding extractor.
PicoAudio adopts a pre-trained VAE model following Liu et al.~\cite{liu2023audioldm1}. 
The diffusion model employs a structure similar to Ghosal et al.~\cite{ghosal2023text} but with fewer parameters, with attention dimensions $\{4, 8, 16, 16\}$, block channels $\{128, 256, 512, 512\}$, and input channels $10$ ($2$ for the timestamp matrix).
HiFi-GAN vocoder is used to transforms spectrogram back to waveform.

PicoAudio is trained for $40$ epochs with a learning rate set to $3\times10^{-5}$ and decreasing according to a linear decay scheduler. 
VAE, LAION-CLAP and HiFi-GAN vocoder are frozen during trainging.
The AdamW optimizer is utilized.
During inference, the Classifier-free guidance scale is set to $3$~\cite{ho2021classifier,nichol2022glide}.

\begin{table*}[t]
\renewcommand{\arraystretch}{1}
    \centering
    \small
    \caption{ 
    Evaluation results.     
    F1$_{\text{segment}}$ / $\bm{L_1^{\text{freq}}}$ respectively measures the \textbf{timestamp alignment} / \textbf{occurrence frequencies error} between generated audio and input conditions.
    \textbf{FAD}  measures the audio quality.
    \textbf{MOS} denotes subjective metrics.
    Ablation study: \textbf{``w/o T"} indicates that the model does not utilize timestamp matrix $\mathcal{O}$, which shares a similar framework with the baseline models.
    }
    \begin{tabular}{c|c|cc|cc|cc|cc}
    \toprule
    \multicolumn{2}{c|}{Condition}&\multicolumn{4}{c|}{Timestamp}&\multicolumn{4}{c}{Occurrence Frequency}  \\
    \midrule
    \multicolumn{2}{c|}{Metrics} &F1$_{\text{segment}}$  &MOS$_{\text{control}}$&FAD$\downarrow$&MOS$_{\text{quality}}$&$L_1^{\text{freq}}\downarrow$&MOS$_{\text{control}}$&FAD$\downarrow$&MOS$_{\text{quality}}$ \\
    \midrule
    \midrule
    \multirow{5}{4em}{Single Event} & Ground Truth   &0.797& 4.78 &0& 4.44 &0.302& 4.9 &0& 4.38 \\
    & AudioLDM2     &0.675& 2.14 &10.853& 3.34    &2.408& 2.3  &20.677& 3.68 \\
    & Amphion       &0.566& 1.98 &11.774&2.82  &2.060& 2.22 &11.999 & 3.54   \\
    & PicoAudio w/o T & 0.694 & 2.78 &  5.926 & 4.2 & 1.25 & 2.92 & 5.923 & 4.2 \\
    & PicoAudio (Ours) &\textbf{0.783}&\textbf{4.58}&\textbf{3.175}&\textbf{4.16}&\textbf{0.537}&\textbf{4.92}&\textbf{2.295}&\textbf{4.1}\\

    \midrule
    \midrule
    \multirow{5}{4em}{Multiple Events} & Ground Truth   &0.787& 4.6 &0& 4.38 &0.447& 4.68 &0& 4.56\\
    & AudioLDM2      &0.593&1.82&10.112&2.36&2.046&2.14&18.334  &2.3  \\
    & Amphion        &0.520&2.2&10.979 &2.72   &1.851&2.48&11.769 &3.24    \\
    & PicoAudio w/o T & 0.614 & 2.12 & 5.218 & 3.42 & 1.216 &2.1 &  5.215 &3.3 \\
    & PicoAudio (Ours) &\textbf{0.772}&\textbf{4.84}&\textbf{2.863}&\textbf{4.12}&\textbf{0.713}&\textbf{4.6}   &\textbf{2.1823}&\textbf{4.38}\\
    \bottomrule
    \end{tabular}
    \label{tab:result}
\end{table*}

\subsection{Evaluation}
Both subjective and objective evaluation metrics are introduced to conduct comprehensive assessments.
\vspace{-2mm}
\paragraph*{Subjective} Mean Opinion Score (MOS) are conducted from two perspectives: audio quality and temporal controllability. 
Audio quality considers the naturalness, distortion, and event accuracy of the generated audio. 
Temporal controllability evaluates the accuracy of timestamp / frequency control.
For each task, $5$ audio clips from each model are rated by $10$ evaluators, and the mean score is calculated. All evaluators are screened for no hearing loss and have university-level education from prestigious universities, using designated headphones.
\vspace{-2mm}
\paragraph*{Objective}
The commonly used FAD in audio generation tasks is utilized to assess the quality of generated audio~\cite{kilgour2019frechet}.
The temporal condition in the timestamp / frequency caption is used as the ground truth for evaluation. 
A grounding model~\cite{xu2024towards} is employed to detect the on- and off-sets of segments in generated audio.
(a) For the timestamp control task, the accuracy of the detected segments is assessed by the segment F1 score~\cite{mesaros2016metrics}, a commonly used metric in sound event detection. 
(b) For the frequency control task, accuracy is measured by the absolute difference between the specified frequency in the caption and the detected frequency in the audio. 
The difference is averaged on test samples $N$ and number of class $C$, denoted as $L_1^{\text{freq}}$:
\begin{equation}
L_1^{\text{freq}} = \frac{1}{N*C}\sum_{n=1}^{N}{\sum_{c=1}^{C}{|\#specified - \#detected|}}
\end{equation}

Simulated audios in the test set are utilized as the ground truth to obtain an objective upper bound, since grounding model cannot detect and localize audio events with $100\%$ accuracy. 

\section{Result}
The control of timestamp and frequency are evaluated separately on both single-event and multiple-event test sets.
The results are presented in Table~\ref{tab:result}.
Two mainstream audio generation models, AudioLDM2~\cite{liu2023audioldm2} and Amphion~\cite{zhang2023amphion,wang2024audit}, are employed as baselines.
Both subjective and objective metrics demonstrate that PicoAudio surpasses baseline models.

\subsection{Timestamp \& Occurrence Frequency Control}
The timestamp controlled audios generated by PicoAudio are very close to the ground truth (upper bound), demonstrating the precision of control, whether in single-event or multi-event tasks.
PicoAudio introduces tailored modules to convert the textual timestamp information into a timestamp matrix, achieving exact control of timestamp in the generated audio at a time resolution of $40$ ms.
Equipped with prompted GPT-4, PicoAudio demonstrates outstanding performance in the frequency error metric $L_1^{\text{freq}}$.
Even in the presence of grounding detection omission errors, it achieves an average error rate of $0.537$ / $0.713$ occurrences per sound event on the single-event / multi-event tasks, respectively.
Achieving $L_1^{\text{freq}}$ less than $1$, akin to the grounding truth, implies that PicoAudio has demonstrated practicality in frequency controlling.

Mainstream generative baseline models, however, fall slightly short in performance.
They obtain lower F1$_{\text{segment}}$ scores and produce a frequency error around $2$ times per event, as they tend to excessively replicate events when faced with temporal conditions.
Furthermore, the ablation study employs a model trained on simulated data without using timestamp matrix $\mathcal{O}$, which shares a similar framework with the baseline models. 
The ablation results lie between the baseline models and PicoAudio, indicating that achieving precise control requires not only temporally-aligned audio-text data but also specific model design.

\subsection{Arbitrary  Temporal Control Capabilities}
With the powerful text processing capabilities of LLM, PicoAudio's precise timestamp control capability provides infinite possibilities for temporal control. 
For instance, for temporal interval and duration control, expressions like \textit{``dog barks three times, with a 2-second interval / duration each time"} can be transformed into single-event timestamp control. 
For events ordering, phrases like \textit{``dog barks then gunshot"} can be transformed into multi-event timestamp control. 
Converting temporal control requirements into timestamp caption format is straightforward for GPT-4 after being prompted.
Therefore, it can be said that \textbf{the PicoAudio can achieve arbitrary precise temporal control}.

However, due to constraints imposed by the audio sources, PicoAudio's limitation lies in its temporary capacity to exercise temporal control over a limited number of events. 
Expanding the quantity of events and achieving comprehensive control beyond temporal are among our future research directions.

\subsection{Audio Quality}
Both the subjective metric MOS$_{\text{quality}}$ and the objective metric FAD demonstrate that PicoAudio outperforms the baseline models.
On one hand, PicoAudio benefits from the advantage of having both the training and test sets derived from simulated data, whereas baseline models have not been trained on such data. 
On the other hand, as mentioned earlier, baseline models tend to excessively replicate events when confronted with temporal control, leading to significant discrepancies with the distribution of the test set.
The ablation experiment demonstrates that solely employing mainstream baseline frameworks trained on simulated data yields limited improvements in audio quality.
Timestamp information aids model in better discerning the distribution of audio.


\section{Conclusion}
Significant progress has been made in audio generation tasks, but performance in terms of temporal control remains subpar, primarily due to the lack of datasets with fine-grained annotations and specific model designs. 
PicoAudio addresses this issue by acquiring data with fine-grained timestamp annotation through web crawling, segmentation, filtering and simulation. 
In terms of model design, PicoAudio utilizes tailored modules to handle temporal information.
It converts captions into one-hot matrices, assisting the diffusion model in achieving $40$ ms level control over timestamp. 
In evaluations encompassing controllability and quality, PicoAudio outperforms mainstream models in both subjective and objective metrics. 
With the support of GPT-4's powerful text processing capabilities, PicoAudio can achieve a variety of temporal control capabilities, including frequency control, interval control, events ordering, etc. 
While PicoAudio's limitation lies in its control over a limited number of events, this serves as a direction for our future work.



\bibliographystyle{IEEEtran}
\bibliography{mybib}

\end{document}